\documentclass{article} 
\usepackage{iclr2025_conference,times}

\usepackage{amsmath,amsfonts,bm}

\def\eqref#1{equation~\ref{#1}}

\def\1{\bm{1}}

\DeclareMathAlphabet{\mathsfit}{\encodingdefault}{\sfdefault}{m}{sl}
\SetMathAlphabet{\mathsfit}{bold}{\encodingdefault}{\sfdefault}{bx}{n}

\usepackage{hyperref}
\usepackage{url}
\usepackage{multirow}
\usepackage{bigfoot}
\usepackage{adjustbox}

\title{Granite Embedding Models}

\author{Granite Team \thanks{See Contributions section for full author list.} \thanks{Please send correspondence to granite-inquiries@ibm.com.}\\
IBM Research AI
}

\iclrfinalcopy 
\begin{document}

\maketitle

\begin{abstract}
We introduce the Granite Embedding models, a family of encoder-based embedding models designed for retrieval tasks, spanning dense-retrieval and sparse-retrieval architectures, with both English and Multilingual capabilities. This report provides the technical details of training these highly effective 12 layer embedding models, along with their efficient 6 layer distilled counterparts. Extensive evaluations show that the models, developed with techniques like retrieval oriented pretraining, contrastive finetuning, knowledge distillation, and model merging significantly outperform publicly available models of similar sizes on both internal IBM retrieval and search tasks, and have equivalent performance on widely-used information retrieval benchmarks, while being trained on high-quality data suitable for enterprise use. We publicly release all our Granite Embedding models under the Apache 2.0 license, allowing both research and commercial use at \url{https://huggingface.co/collections/ibm-granite}.
\end{abstract}

\begin{figure}[h]
    \centering
    \includegraphics[width=1.0\linewidth]{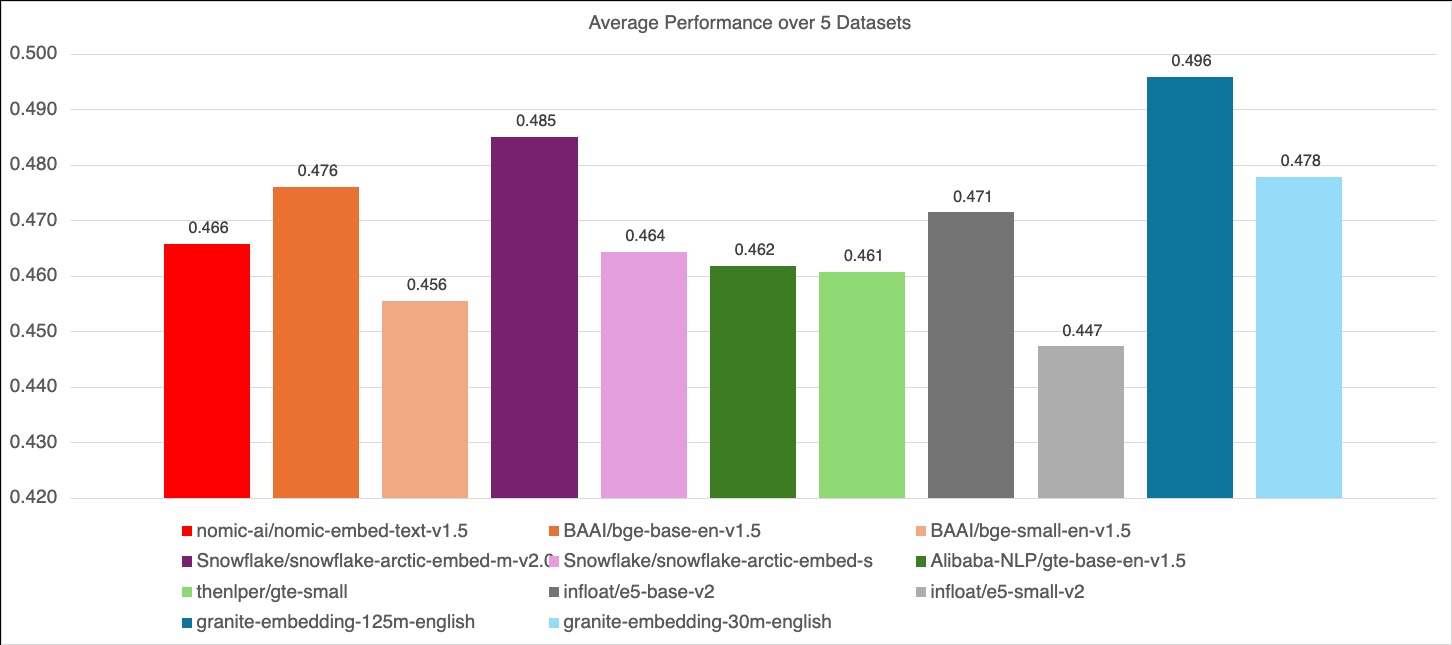}
    \caption{Average performance on the Granite embedding models (in blue) vs BGE, GTE, Snowflake, E5, and Nomic models on 5 QA and IR datasets: BEIR, ClapNQ, CoIR, RedHat, and UnifiedSearch (the last 2 are internal IBM datasets).}
    \label{fig:result-graph}
\end{figure}

\section{Introduction}

The goal of text embedding models is to convert variable length text into a fixed vector, encoding the text semantics into a multidimensional vector in such a way that semantically close texts are close in the vector space, while dissimilar texts have a low similarity. These embeddings can then be used in a variety of tasks, most commonly in retrieval applications, where the relevance of a document to a given query can be determined by the similarity of their embeddings \citep{dunn2017searchqa, xiong2020approximatenearestneighbornegative, neelakantan2022textcodeembeddingscontrastive}\citep{10.1145/3269206.3271800, zhao2020spartaefficientopendomainquestion}, but also in document clustering \citep{angelov20} and text classification \citep{sun2019fine}.

While Large Language Models (LLMs) have been shown to be powerful sources for text embeddings \citep{lee2024nvembed, wang2023e5mistral}, inference with these models are often impractical in an industrial setting, due to large latency and memory footprint. Encoder-based text embedding models \citep{wang2022e5, bge_embedding, chen2024bgem3, merrick2024arcticembedv1} are thus more popular, however they are often trained on data with non-commercial licenses, leading to restrictions in commercial deployment. 

This report introduces Granite Embedding models, a family of encoder embedding models purpose-built for retrieval tasks. All Granite Embedding models have been trained on high quality, curated data, with data quality checks and screening to remove personal information and profane language. We release these models under the Apache 2.0 license, permissible for both commercial and research applications. We release 5 models of varying language support, sizes, and vector types\footnote{The main difference between the sizes of the models comes from the vocabulary: the multilingual models have a large vocabulary of 250K tokens, while the English models have 50k token vocabulary (Table \ref{table:architecture}); the networks themselves have the sane number of parameters. We have reused the RoBERTa \citep{liu2019roberta} and XLM-RoBERTa \citep{conneau-etal-2020-unsupervised} vocabularies in these models.}:
\begin{itemize}
    \item Dense English Models: 30M and 125M parameter models, specialized for English-only retrieval applications
    \item Dense Multilingual Models: 107M and 278M parameter models, finetuned for multilingual retrieval for 12 languages
    \item Sparse English Model: a 30M model adapted for sparse retrieval capabilities in English
\end{itemize}

The granite-embedding models are specifically adapted for enterprise usage, offering strong performance for a variety of inference budgets. The base-size models (12 layers) have strong performance on both academic and enterprise tasks, and the light-weight 6-layer models provide good performance with very low latency and memory requirements. 

The structure of the paper is as follows: Section \ref{sec:arch} describes the model architecture and training background, the training data and its curation is described in Section \ref{sec:data}. Section \ref{sec:dense-english} describes in detail the training recipes for the Granite Dense English Models, while Section \ref{sec:dense-multi} describes the dense multilingual models and Section \ref{sec:sparse-english} describes the sparse English model. Finally, Section \ref{sec:evaluation} presents a comprehensive evaluations of the granite embedding models, comparing their performance with other open source encoder embedding models.

\section{Granite Embedding Models}
\label{sec:arch}
The Granite Embedding collection delivers innovative sentence-transformer models purpose-built for retrieval-based applications. Featuring a bi-encoder architecture \citep{reimers-gurevych-2019-sentence}, these models generate high-quality, low-dimensional embeddings for textual inputs -- such as queries, passages, and documents -- enabling seamless comparison through cosine similarity. In this Section, we describe the architectural details of our models, along with a general overview of our training methodology.

\paragraph{Model Architecture}
The Granite Embedding family consists of five encoder-only models of varying sizes and vocabularies for a diverse set of use-cases, featuring English and Multilingual models targeted for retrieval applications. The English models are RoBERTa-like \citep{liu2019roberta} models while the Multilingual models follow an  XLM-RoBERTa \citep{conneau-etal-2020-unsupervised} configuration, all trained internally at IBM Research. The Multilingual models are targeted towards 12 languages, however, they can be finetuned on other languages covered the XLM-RoBERTa vocabulary (approximatively 100 languages).

We release five high performing base-sized models, {\em granite-embedding-125m-english} and {\em granite-embedding-278m-multilingual}, as well as smaller, faster counterparts, {\em granite-embedding-30m-english}, {\em granite-embedding-107m-multilingual} and {\em granite-embedding-30m-sparse} suitable for diverse resource and latency budgets.   

Dense Granite embeddings models use the final hidden state of the CLS token, or $<$$s$$>$ for Roberta-based models, as the embedding representation (also known as CLS pooling), which we find to perform better than mean-pooling strategy commonly used in other works \citep{wang2022e5, li2023gte, nussbaum2024nomic}. The sparse Granite embedding model uses max-pooling across all sequence tokens to produce a variable length weighted bag-of-words like output. The maximum context length of all Granite embedding models is 512 tokens, and longer sequences are truncated to this length.

Detailed specifications of the architecture of each model is shown in Table~\ref{table:architecture}.

\begin{table}[!ht]
    \label{table:architecture}
    \centering
    \setlength{\tabcolsep}{0.5em} 
    {\renewcommand{\arraystretch}{1.3}
    \begin{tabular}{r|cc|cc}
    \hline
        Model & 30M & 125M & 107M & 278M \\ \hline
        Embedding size & $384$ & $768$ & $384$ & $768$ \\ 
        Layers & $6$ & $12$ & $6$ & $12$ \\  
        Intermediate size & $1536$ & $3072$ & $1536$ & $3072$ \\
        Max. Sequence Length & $512$ & $512$ & $512$ & $512$ \\ 
        Vocabulary Size & \multicolumn{2}{c|}{$50265$}  & \multicolumn{2}{c}{$250002$}  \\ 
        \hline
        \multirow{2}{*}{Target Languages} & \multicolumn{2}{c|}{\multirow{2}{*}{En}} & \multicolumn{2}{c}{En, Ar, Cs, De, Es, Fr,}\\
        & \multicolumn{2}{c|}{} &\multicolumn{2}{c}{It, Ja, Ko, Nl, Pt, Zh}\\
        \hline
    \end{tabular}}
    \caption{Architectural Details for Granite Embedding Models}
\end{table}

\paragraph{Training Recipe}
Embedding models are trained using a contrastive learning objective \citep{gao-etal-2021-simcse}, which brings the embeddings of a query closer to those of relevant passages, and pushes them further away from non-relevant ones. In this work, we use an improved contrastive loss proposed in \citet{li2023gte} which introduces an additional bidirectional signal to expand negatives. Specifically, for a batch of triples $([q_i,(p_{ij})_j])_i$ consisting of a query and a set of passages -- without loss of generality, we can assume that $p_{i0}$ is a positive passage for query $i$, while $p_{ij}, j>0$ are negative passages -- we define the contrastive loss as:

\begin{equation}
    \mathcal{L}_{C} = -\frac{1}{n} \sum_{i=1}^{n} \mathrm{log} \frac{e^{s(q_i, p_{i0})}}{Z_i}
\label{eq:cont-loss}
\end{equation}
\begin{equation}
\begin{aligned}
Z_i = e^{s(q_i, p_{i0})} + \alpha\sum_{j>0} e^{s(q_i,p_{ij})} + \beta\sum_{i' \neq i} e^{s(q_i,q_{i'})} + \gamma\sum_{j>0} e^{s(p_{i0},p_{ij})}
\label{eq:z-comp}
\end{aligned}
\end{equation}

where $s(q, p)$ is a the temperature-scaled cosine similarity between the embeddings of $q$ and $p$:
\begin{equation}
    s(q,p) = \frac{1}{\tau} \frac{\mathbf{E}(q) \cdot\mathbf{E}(p)} {\|\mathbf{E}(q)\|\|\mathbf{E}(p)\|}
\label{eq:sim-score}
\end{equation}

where $\mathbf{E}(p)$ is the embedding of text $p$\footnote{Note that, for high temperatures $\tau$, this loss is a smooth version of maximizing \[s(q_i,p_{i0})-\max\left( \alpha\max_{j>0}s(q_i,p_{ij}), \beta\max_{i'\neq i}s(q_i,q_{i'}), \gamma\max_{j>0}s(p_{i0},p_{ij})\right)\]}, here $\alpha, \beta$, and $\gamma$ are added so that we can remove any of these components easily (they can be either $0$ or $1$). Contrastive training is conducted using paired data sourced or generated from a diverse set of domains. Similar to prior work \citep{wang2022e5, merrick2024arcticembedv1, bge_embedding, nussbaum2024nomic} we use both weakly-paired data from web sources, along with high quality annotated data and targeted synthetic data. We use in-batch negatives to better approximate the contrastive signal, and also include hard negatives for a subset of our data consisting of high-quality annotated pairs, mining them whenever needed.

We explore additional techniques to improve the performance of our models. For our English models, we find it helpful to include an initial \textbf{retrieval-oriented pretraining} step, similar to \citet{xiao-etal-2022-retromae}, which uses an asymmetric model structure and masked auto-encoding to improve embedding quality. We also find that \textbf{knowledge distillation} techniques, particularly distilling the distribution of similarity scores from larger models, are extremely effective in improving the performance of our models: {\em granite-embedding-125m-englis} is distilled from a larger decoder-based embedding model, {\em granite-embedding-278m-multilingual} uses self-distillation to improve performance and all our small models are distilled from the larger encoder-based embedding models. Finally, we use \textbf{model merging} \cite{xiao2023lmcocktailresilienttuninglanguage} to help adapt our models to specific domains without losing general performance. More details on training are provided in the subsequent sections.

\section{Training Data}
\label{sec:data}

Granite Embedding Models are trained on two types of data: 
\begin{enumerate}
    \item weakly paired data mined from the web with in-batch negatives and
    \item high quality, annotated data with hard negatives for finetuning derived from three key sources: 
    \begin{enumerate}
    \item publicly available paired data, 
    \item IBM-internal paired data targetting specific technical domains, and
    \item IBM-generated synthetic data. 
    \end{enumerate}
\end{enumerate}
For governance, our data undergoes a data clearance process subject to technical, business, and governance review. This comprehensive process captures critical information about the data, including, but not limited to, its content description, intended use, data classification, licensing information, usage restrictions, as well as an assessment of sensitive information (i.e, personal information). 

\paragraph{Weakly Paired Data with In-Batch Negatives} The bulk of this data consists of title-body pairs obtained from diverse web sources such as Wikipedia, StackExchange, Semantic Scholar, Arxiv and PubMed. We also include citation pairs (SPECTER \citep{specter2020cohan}, Semantic Scholar \citep{lo-etal-2020-s2orc}), duplicate questions (StackExchange), and question answer pairs (PAQ \citep{lewis_paq_2021}, WikiAnswers 
 \citep{fader_2014_wikianswers}, SearchQA \citep{dunn2017searchqa}). For multilingual models, a subset of the English data is used, along with title-body pairs from mC4, multilingual Wikipedia, and multilingual Webhose, and  Machine Translations of StackExchange. While training with these larger datasets, we only include in-batch negatives to provide a contrastive signal.

\paragraph{Annotated Data with Hard Negatives} We use high-quality open source question-answer pairs from NQ \citep{kwiatkowski-etal-2019-natural-questions}, SQuAD \citep{squad2} and HotpotQA \citep{yang-etal-2018-hotpotqa}, and fact verification pairs from FEVER \citep{thorne-etal-2018-fever}. Notably, while other open-source models trained on the popular MS-MARCO retrieval dataset \citep{bajaj2018msmarco}, we chose not to train on it given it is designed for research purposes only rather than commercial use. For our multilingual models, we also include data from MIRACL \citep{zhang-etal-2023-miracl}, TyDiQA \citep{clark-etal-2020-tydi} and Sadeem Question Answering \citep{sadeem_qna}. Additionally, we find it helpful to include cross-lingual, translated parallel corpora to help improve multilingual performance. For enterprise applications, we create title-body pairs from IBM internal datasets targeting specific technical domains. For all these datasets, we mine additional hard negatives for a stronger contrastive signal (except MIRACL and TyDi, which have annotated hard negatives). 

\paragraph{Synthetic Data} In addition to the above sources, we generate high quality synthetic multilingual queries with LLMs\footnote{\verb|mistralai/Mixtral-8x22B-Instruct-v0.1| ,  \verb|microsoft/Phi-3.5-MoE-instruct|} \citep{jiang2024mixtralexperts, abdin2024phi3technicalreport} for Wikipedia paragraphs in all our target languages. We carefully design prompts that generate a diverse set of queries for a given input paragraph, such as simple keyword search queries, queries that can be partially/completely answerable from the given paragraph, facts that are validated by the paragraphs, summaries of the paragraph, etc. in order to help the model learn different notions of similarity. We also generate hard negatives for these queries, where we instruct the model to modify the positive passage such that it maintains the context of it, but does not answer the given query.

\paragraph{Hard Negative Mining} For high quality datasets that are not annotated with hard negatives, we mine them using a pre-existing lightweight embedding model trained at IBM Research. We calculate the similarity of non-relevant passages from the same dataset with the given query, and sample negatives from the top-100  ranked passages. We filter out false negatives by discarding passages whose similarity to the positive passage lies above a threshold. For fine tuning, we use a single hard negative, along with random in-batch negatives for the annotated data, and find that increasing the number of hard negatives does not significantly help performance. For distillation, we use up to three hard negatives and find out it improves the performance over single hard negative. We suspect that this is because more hard negatives fill the gap between hard negative similarity and in-batch negative similarity. Therefore provides more signal for distillation.

\paragraph{Data Sampling} During training, each batch is sourced from a single dataset. To address the imbalance between the sizes of the datasets used for contrastive training, we sample batches from each dataset $D_i$ proportionately to its size, $|D_i|$, where the sampling probability is defined by:
\begin{equation}
    p_i = \frac{{|D_i|}^\alpha}{\sum_{i}{{|D_i|}^\alpha}}
\end{equation}
where $\alpha$ is a hyperparameter. We observe $\alpha$ value of 0.5 and 0.9 to result in best scores for English and multilingual models resp.

\section{Dense Retrievers: English}
\label{sec:dense-english}

\subsection{granite-embedding-125m-english}
\label{subsec: granite-125m}

{\em granite-embedding-125m-english} is a dense bi-encoder embedding model that has the same architecture as RoBERTa-base \citep{liu2019roberta}. It performs well on many enterprise use cases while maintaining competitive scores on academic benchmarks such as BEIR. The starting point of fine tuning this model is a RoBERTa-base like model, called WatBERT, internally trained by IBM using carefully curated data. The model is then further fine tuned through two steps: (1) retrieval oriented pretraining, (2) knowledge distillation from decoder based embedding model. 

\paragraph {Retrieval Oriented Pretraining} \label{RetroMAE}  We first apply retrieval oriented pretraining to the WatBERT model similar to Retro-MAE \citep{xiao-etal-2022-retromae}. The training follows a masked auto-encoding workflow. The input sentence is masked twice with different ratios. One moderately masked input (15\%-30\%) is used by the encoder model where the sentence embedding is generated from the [CLS] token. The sentence embedding joins with the other aggressively masked input (50\%-70\%) and is used by a single-layer transformer to recover the original sentence via masked language modeling. We train the model with 56 million sentences from Wikipedia, BookCorpus and StackExchange.

\paragraph{Decoder based teacher model} We further fine tune the Mistral-7B-Instruct-v0.2 \citep{jiang2023mistral7b} using contrastive training as our teacher model. Similar to \citep{wang2024improvingtextembeddingslarge}, for each query-document pair (q, p), we apply an instruction template to the original query to generate a new one:
\begin{equation}
    q_{inst} = Instruct: \{task\_definition\} Query: {q}
\end{equation}
where "\{task\_definition\}" is a placeholder for a one-sentence description of the embedding task. We then append an [EOS] token to the end of the query and document and feed them into the Mistral model to obtain the query and document embeddings by taking the [EOS] vector of the last layer. 

We train two separate embedding models from the Mistral-7B-Instruct-v0.2 and create the final teacher model by merging the two models together. The first model is trained using three million pairs of weakly paired data with in-batch negatives. The second model is trained using one million annotated data with one hard negative. 

\paragraph{Distill from decoder based model} Using the same methodology as described in Section  \ref{subsec: contrastive-dist}, we then distill the knowledge of mistral-7b-instruct-v0.2 embedding model into {\em granite-embedding-125m-english}. Notably, the similarity score of the decoder model is calculated from the [EOS] token while the similarity score of the encoder model is calculated from the [CLS] token. The distillation also happens in two stages. In the first stage, weakly paired data with in-batch negative is used while in the second stage, high quality annotated data with three hard negatives is used. 

\subsection{Granite-Embedding-30M-English}
\label{subsec: granite-30m-dense}
{\em granite-embedding-30m-english} is an efficient embedding model distilled from an independently trained encoder model similar to {\em granite-embedding-125m-english}. This model is about 4 times faster than  {\em granite-embedding-125m-english}, while maintaining good performance, and is highly suitable to use in latency-constrained applications. This model is trained in three steps: (1) Retrieval Oriented Pretraining, 2) contrastive based distillation on a large corpus of data (3) targeted enterprise-domain adaption.

\paragraph{Retrieval Oriented Pretraining} 
 We prime our model with RetroMAE Based Distillation. First, we incorporate supervision from the RetroMAE-pretrained encoder described in \ref{RetroMAE} using distillation. We substitute the standard masked language modeling loss for the encoder with a distillation loss between the student's and teacher's output distribution, while keeping the unmodified decoder loss from \citet{xiao-etal-2022-retromae}. We use Wikipedia, BookCorpus and StackExchange in this step.

\paragraph{Contrastive Based Distillation} 
\label{subsec: contrastive-dist}
We tune the embeddings by distilling the distribution of similarity scores of the teacher model. Specifically, using the same notations from Section \ref{sec:arch}, the training objective minimizes the cross entropy between the teacher's distribution $P_t$ of similarity scores between pairs and the student's distribution, $P_s$. Following \citet{hinton2014distilling}, we also scaled the score distribution of both teacher and student by a temperature, $\tau$:
\begin{equation}
\label{eq:kd-loss}
    \mathcal{L}_{KD} = - \sum_{i=1}^{n} \sum_{j=1}^{m} P_t(q_i, p_{i0}) \mathrm{log} P_s(q_i, p_{i0})
\end{equation}
\begin{equation}
    P_s(q_i, p_{i0}) =  \frac{e^{s_s(q_i, p_{i0})/\tau}}{Z_{s,i}}
\end{equation}
\begin{equation}
    p_t(q_i, p_{i0}) = \frac{e^{s_t(q_i, p_{i0})/\tau}}{Z_{t,i}}
\end{equation}

Here, $s_s(q_i, p_{i0})$ and $s_t(q_i, p_{i0})$ represent the similarity scores between questions and reference passages $\{q_i, p_{i0}\}_i$ defined in Equation \ref{eq:sim-score}, for the student and teacher respectively, and $Z_{s,i}$ and $Z_{t,i}$ are computed as in Equation \ref {eq:z-comp}, with $\beta=0$ and $\gamma=0$.

Utilizing the similarity scores for distillation instead of the distilling the text embedding vectors directly allows for effective knowledge transfer between models of different embedding sizes, without introducing additional parameters for embedding projection. For datasets without a mined or annotated hard negative, we find that perturbing the positive passage slightly (through deletion or swapping of spans of text) is an effective and inexpensive way to create a richer distribution of similarity scores, leading to better performance after distillation. While the resulting passage may not necessarily be a \emph{true negative}, the lack of an explicit contrastive signal differentiating between positive and negative passages during score distillation allows this to be an effective way of using a large quantity of unsupervised data. 

\paragraph{Targeted Enterprise-Domain Adaption} In order to strengthen performance on specific enterprise domains, we conduct another round of distillation on IBM-internal paired data. We use model merging \citep{xiao2023lmcocktailresilienttuninglanguage} between this domain-adapted model and the model trained on all the data, resulting in the best overall performance in both enterprise-specific and general-purpose domains.

\section{Dense Retrievers: Multilingual}
\label{sec:dense-multi}

We develop multilingual variants of Granite Embedding models, targeted towards 12 languages, trained on a large amount of multilingual data including synthetically generated pairs. These models are trained by gradually expanding the number of languages in the training data: we start with older, un-released versions of these models trained on only 6 languages (English, French, German, Japanese, Portuguese, and Spanish), and further train these models on annotated or synthetic data from all our target languages (aforementioned languages along with Arabic, Czech, Dutch, Italian, Korean and Chinese). We find that this helps maintain the performance of the original models on the languages they are trained on, while improving performance on new languages added in the mix.  For the multilingual models, a retrieval-oriented pretraining step does not improve the model's performance after finetuning, so we omit this initial step.

\subsection{Granite-Embedding-278M-Multilingual}

{\em granite-embedding-278m-multilingual} is an effective, base-size model specifically trained for retrieval applications in multilingual text settings for 12 Languages. To build this model, we start from an in house XLM-RoBERTa like model, trained on 100 languages, and first train it for retrieval on a 6-languages using a stage-wise contrastive training approach, common in previous works. This is followed by extending finetuning on all our target langauges. Finally, we conduct a self-distillation step, which we find to be extremely useful in improving the performance of our model without additional data or an external model. 

\paragraph{Stage-wise Contrastive Training} First, we train a six-language model using a stage-wise training approach, as done in prior work \citep{bge_embedding, chen2024bgem3, wang2022e5, li2023gte, zhang2024mgte}. Specifically, we start with an \emph{unsupervised pretraining} phase, where we train on a large corpus of naturally occuring pairs (eg. title-body) mined from web sources. Here we use a large batch size and in-batch negatives for effective learning. We then conduct a \emph{supervised finetuning} step using annotated pairs with a single hard negative. This results in a model with good performance on English, French, German, Japanese, Portuguese, and Spanish.

\paragraph{Target Language Finetuning} To expand the scope of this model, we conduct further finetuning on all our 12 target languages. This is done with annotated and synthetic paired data with a single hard negative. We find this strategy to be more effective at maintaining prior performance than training a 12-language model from scratch.

\paragraph{Self Distillation} We find that an additional training step further improves the performance of this model. Self Knowledge Distillation is a technique where the knowledge of a model is transferred into another model of the same size using its temperature-scaled output distribution. Prior work has shown this technique to be an effective method of progressive performance improvement \citep{BornAgainNetworks, Kim_2021_selfkd}, and we interpret this technique as a way to reduce the effect of any noisy labels. Specifically, we distil the output score distribution of the finetuned 12-language model into a student initialized from the same teacher on the distillation loss of Equation \ref{eq:kd-loss}. We use all the finetuning data available for this distillation step, including our synthetic data and enterprise-domain data.

\subsection{Granite-Embedding-107M-Multilingual}

{\em granite-embedding-107m-multilingual} is an efficient multilingual model, almost 4 times faster than {\em granite-embedding-278m-multilingual} while maintaining good performance. This model is distilled from components of the {\em granite-embedding-278m-multilingual} model: first, we distill the intermediate 6-language base model using data in these 6 languages, then continue distilling from the 12-layer model on all languages.

Like {\em granite-embedding-30m-english}, this model is trained by distilling with all the available data, using a single hard negative when available, and perturbing the positive to get a richer probability distribution for datasets without hard negatives. We also use in-batch negatives to better approximate the contrastive signal.

Similar to it's larger counterpart, {\em granite-embedding-107m-multilingual} is constructed by first training a 6-language model, by distilling from the intermediate base model on all the training data for English, French, German, Japanese, Portuguese, and Spanish. Then, we distill from {\em granite-embedding-107m-multilingual} on all the 12 target languages. Distilling in stages yields a better overall performance than distilling all 12 languages in a single step. 

\section{Sparse Retrievers: English}
\label{sec:sparse-english}

The Granite dense retriever models, described in \ref{sec:dense-english} , produce a fixed-size embedding vector for a query or document. Alternatively, Granite sparse retriever model is trained to produce a variable length bag-of-word like dictionary, containing expansions of sentence tokens and their corresponding weights. This sparse representation could be useful for exact matching of terms and could be efficiently ingested as an inverted index as shown by \cite{10.1145/3269206.3271800, bai2020spartermlearningtermbasedsparse, dai2019contextawaresentencepassagetermimportance,  formal2021spladesparselexicalexpansion}. 

\subsection{Granite-Embedding-30m-Sparse}

{\em granite-embedding-30m-sparse} is a 6 layer model, similar in architecture to {\em granite-embedding-30m-english}, and is trained following the first two training steps of the dense counterpart \ref{subsec: granite-30m-dense}:(1) Retrieval Oriented Pretraining (2) contrastive-based distillation on a large corpus of data. 

Sparse models are popularly trained to predict the importance of each vocabulary term based on the logits of masked-language model head \citep{bai2020spartermlearningtermbasedsparse, formal2021spladesparselexicalexpansion}. We start with a 6 layer WatBERT model, apply Retro-MAE based distillation, and then further train the language model head using contrastive-based knowledge distillation from a larger (125m) dense retriever teacher. We experiment with Margin-MSE loss \citep{hofstätter2021improvingefficientneuralranking}, but find Cross Entropy to be more effective in our setup, possibly because our teacher is a dense retriever. Further, we introduce another loss, as described below, to train our sparse model.

\paragraph{Term Weight Pooling Strategy} We follow \cite{formal2021spladev2sparselexical} and use max-pooling to compute the importance $w_{j}$ for each vocabulary token $j$ for input sequence $I$:

\begin{equation}
    w_j = \max_{i\in I}  \log(1 + \text{ReLU}(w_{ij}))
\end{equation}

\paragraph{Learning Sparse Representations} We observe that minimizing the FLOPS loss \cite{paria2020minimizingflopslearnefficient}, as shown in SPLADE \citep{formal2021spladesparselexicalexpansion, formal2021spladev2sparselexical}, did not lead to a sufficiently sparse model. We hypothesize this being an artifact of our starting point, the 6 layer WatBERT model, as the 6 layer WatBERT model is distilled from the 12 layer WatBERT \ref{subsec: granite-125m} using multi-head self-attention relation distillation \citep{wang2021minilmv2multiheadselfattentionrelation} which does not train the Masked Language Modeling head. So, in addition to the FLOPS loss we introduce a NORM loss that learns to minimize the total norm of the sentence with the intuition being this would further bring the term weight close to zero, as follows:

\begin{equation}
\label{eq:norm-loss}
    \mathcal{L}_{NORM} = \sum_{j} \max_{i\in I}  \log(1 + \text{ReLU}(w_{ij}))
\end{equation}

\paragraph{ Overall Loss } We jointly optimize the sparse model with contrastive distillation \ref{eq:kd-loss}, FLOPS regularization, and NORM regularization  \ref{eq:norm-loss} losses: 

\begin{equation}
    \mathcal{L} = \mathcal{L}_{KD} + \lambda_q \mathcal{L}_{FLOPS}^q + \lambda_p \mathcal{L}_{FLOPS}^p + \sigma_q \mathcal{L}_{NORM}^q + \sigma_p \mathcal{L}_{NORM}^p 
\end{equation}

where $\lambda_q$ and $\lambda_p$ are FLOPS regularization weights and $\sigma_q$ and $\sigma_p$ are total norm regularization weights for query and passage resp.

\section{Evaluation}
\label{sec:evaluation}

We evaluate the performance of the Granite Embedding models on a variety of retrieval tasks, spanning multiple domains and languages. Figure \ref{fig:result-graph} shows comparative performance of the english Granite Embedding models with other open source models.  Results show the Granite Embedding models achieve higher scores in average than the all other models. In this section we show results that further show that the Granite Embedding models are faster, achieve higher average accuracy over a variety of benchmarks, and that they not only maintain competitive scores on benchmarks such as MTEB, but also out-perform the other models on Code Retrieval. 

We compare our encoders with other state-of-the-art embedding models of similar parameter size. Specifically, we compare {\em granite-embedding-125m-english} to the BGE Base \citep{bge_embedding} and E5 Base Models \citep{wang2022e5}, which have identical architecture as the granite base models, with a smaller vocabulary size. {\em granite-embedding-30m-english} is compared to the small versions of the BGE and E5 model family: note that these models have double the number of layers than the Granite 30M models. We compare the multilingual models against the Multilingual E5 \cite{wang2024multilinguale5}  Base and Small models respectively. For sparse embeddings, we compare {\em granite-embedding-30m-sparse} with SPLADE-v3-DistilBERT \cite{lassance2024spladev3newbaselinessplade}, a 6-layer model with larger 768 embedding size.

\subsection{English Retrieval Performance}
All Granite Embedding models are evaluated on retrieval tasks from the MTEB benchmark \citep{muennighoff2022mteb}, which consists of datasets and evaluation from BEIR \citep{thakur2021beir}. This consist of 15 datasets spanning multiple domains, and is used to test the model's ability to find the relevant document for a given query.  The experiment results are reported in Table \ref{results:beir}. We report the nDCG@10 scores on each dataset, showing strong performance relative to other open-source models of similar sizes. We also note the average performance on all tasks except MS-MARCO retrieval \cite{bajaj2018msmarco}, for a fair zero-shot comparison, as other embedding models train on MS-MARCO, but our models do not, due to unfavorable license. Despite being trained on less data, and only permissibly licensed public datasets, our models show strong performance.

\begin{table}[t!]
\centering
\begin{adjustbox}{width=1\textwidth}
\setlength{\tabcolsep}{0.5em} 
{\renewcommand{\arraystretch}{1.3}
\begin{tabular}{p{3.2cm}|p{0.8cm}p{0.9cm}p{0.8cm}p{0.8cm}p{0.8cm}p{0.8cm}p{0.8cm}p{1cm}p{0.8cm}p{0.8cm}p{0.8cm}p{0.8cm}p{0.8cm}p{0.8cm}p{0.8cm}|p{0.7cm}p{1cm}}
\hline
Model & Argua-Ana & Climate Fever & CQA-Dup-stack & DB-Pedia & Fever & FiQA & HotPot-QA & MS-Marco & NF-Corpus & NQ & Quora & Sci-Fact & Sci-docs & Touche & Trec-Covid & \textbf{Avg.} & \textbf{Avg. (no MSMarco)} \\
\hline
e5-small-v2 & 41.7 & 22.9 & 37.1 & 41.3 & 81.6 & 37.4 & 66.6 & 41.5 & 32.5 & 59.1 & 85.7 & 68.9 & 17.8 & 27.1 & 74.5 & 49.0 & 49.6 \\
bge-small-en-v1.5 & 59.6 & 31.8 & 39.1 & 40.0 & 86.6 & 40.3 & 69.9 & 40.8 & 34.3 & 50.2 & 88.8 & 71.3 & 20.5 & 26.0 & 75.9 & 51.7 & 52.5 \\
e5-base-v2 & 44.5 & 26.6 & 38.5 & 42.2 & 85.0 & 39.9 & 69.2 & 41.8 & 35.4 & 58.2 & 86.6 & 71.9 & 18.7 & 26.4 & 69.6 & 50.3 & 50.9 \\
bge-base-en-v1.5 & 63.6 & 31.2 & 42.4 & 40.8 & 86.3 & 40.7 & 72.6 & 41.4 & 37.4 & 54.2 & 88.9 & 74.0 & 21.7 & 25.7 & 78.1 & 53.3 & 54.1 \\
\textbf{granite-embedding-30m-english} & 56.4 & 30.3 & 44.3 & 36.0 & 85.5 & 36.9 & 62.9 & 30.7 & 33.7 & 51.6 & 86.7 & 71.3 & 22.5 & 24.0 & 63.1 & 49.1 & 50.4 \\
\textbf{granite-embedding-125m-english} & 58.4 & 33.2 & 48.2 & 39.4 & 88.2 & 44.9 & 67.8 & 32.5 & 37.3 & 58.0 & 87.8 & 74.7 & 24.2 & 20.2 & 69.3 & 52.3 & 53.7 \\
\hline
\hline
multilingual-e5-small & 39.1 & 22.6 & 36.1 & 37.8 & 75.3 & 33.3 & 65.1 & 41.0 & 31.0 & 56.3 & 86.9 & 67.7 & 13.9 & 21.2 & 72.6 & 46.6 & 47.0 \\
multilingual-e5-base & 44.2 & 23.9 & 38.5 & 40.4 & 79.4 & 38.2 & 68.6 & 42.3 & 32.5 & 60.0 & 87.7 & 69.3 & 17.2 & 21.4 & 69.8 & 48.9 & 49.4 \\
\textbf{granite-embedding-107m-multilingual} & 51.7 & 25 & 39.9 & 32.8 & 83.2 & 28.4 & 58.8 & 28.3 & 27 & 50.9 & 85.8 & 63.3 & 17.4 & 24.9 & 62.8 & 45.3 & 46.6 \\
\textbf{granite-embedding-278m-multilingual} & 55.2 & 29.5 & 42.7 & 34.6 & 84.9 & 35.7 & 61.5 & 29.9 & 28.9 & 53.4 & 86.9 & 17.7 & 65.9 & 27.3 & 68.6 & 48.2 & 49.5 \\
\hline
\hline
splade-v3-distilbert & 48.4	& 22.8 & - &		42.6 &	79.6&	33.9	&67.8	& - & 34.8	&54.9	&81.7	&14.8	&68.5&	30.1&	70 &	50.0$^{\dagger}$  &	50.0*\\
\textbf{granite-embedding-30m-sparse} & 54.7	&25.5&	42.1 & 37.2&	82.0&	35.2&	68.1&	30.3&	32.3&	50.8&	85.1&	71.4&	21.3&	22.8&	71.9&	48.7&	50.6*\\
\hline
\end{tabular}}
\end{adjustbox}
\caption{BEIR benchmark \citep{thakur2021beir} nDCG@10 scores. $\dagger$ Average is not directly comparable as splade-v3-distilbert \citep{lassance2024spladev3newbaselinessplade} does not report BEIR scores on CQADupStack and MS-MARCO. * Average does not include CQADupstack.}
\label{results:beir}
\end{table}

While our models are purpose-built for retrieval applications, we also evaluate them on the entire MTEB benchmark (including performance on non-retrieval tasks such as classification, clustering, pair clustering, re-ranking, retrieval, sentence similarity and summarization) in  Appendix \ref{app:eng-mteb} to measure the quality of embeddings for non-retrieval tasks.

\subsection{Multilingual Retrieval Performance}

The Multilingual Granite Embedding models are evaluated on the Miracl Benchmark \citep{zhang-etal-2023-miracl} and the Mintaka Retrieval task \citep{sen-etal-2022-mintaka}, both of which measure the performance of monolingual retrieval on multiple languages. Miracl is a high-quality text retrieval benchmark sourced from Wikipedia, in 18 langauges. Mintaka Retrieval comprises of Wikipedia-based questions translated into 8 langauges. We report the nDCG@10 scores for these models on both of these tasks in Table \ref{results:miracl} and Table \ref{results:mintaka}.

\begin{table}[ht!]
\centering
\begin{adjustbox}{width=1\textwidth}
\setlength{\tabcolsep}{0.5em} 
{\renewcommand{\arraystretch}{1.3}
\begin{tabular}{p{3.1cm}|llllllllllllllllll|l}
\hline
Model & ar & bn & de & en & es & fa & fi & fr & hi & id & ja & ko & ru & sw & te & th & yo & zh & \textbf{Avg}. \\
\hline
multilingual-e5-small & 71.4 & 68.3 & 48.8 & 48 & 50.8 & 53.3 & 73.3 & 47.1 & 55.1 & 50.7 & 63.6 & 61.2 & 59 & 68.5 & 81.3 & 74.9 & 45.3 & 45.9 & 59.3 \\
multilingual-e5-base & 71.6 & 70.2 & 52.1 & 51.2 & 51.5 & 57.5 & 74.3 & 49.7 & 58.3 & 51 & 64.7 & 62.3 & 61.6 & 71.1 & 75.1 & 75.3 & 70.6 & 57.5 & 62.5 \\
\textbf{granite-embedding-107m-multilingual} & 62.5 & 65.5 & 45.1 & 46.7 & 46.6 & 50 & 65.6 & 47.6 & 42.1 & 45.5 & 59.2 & 58.4 & 48.3 & 59.1 & 78.2 & 71.8 & 64.7 & 48.4 & 55.9 \\
\textbf{granite-embedding-278m-multilingual} & 64.2 & 68.1 & 48.1 & 49.4 & 49.7 & 50.2 & 67.5 & 49.9 & 46.1 & 47.2 & 62.8 & 59.2 & 52.3 & 61.3 & 79.2 & 73.3 & 68.7 & 52.6 & 58.3 \\
\hline
\end{tabular}}
\end{adjustbox}
\caption{MIRACL benchmark \citep{zhang-etal-2023-miracl} nDCG@10 scores.}
\label{results:miracl}
\end{table}

\begin{table}[ht!]
\centering
\begin{adjustbox}{width=1\textwidth}
\setlength{\tabcolsep}{0.5em} 
{\renewcommand{\arraystretch}{1.3}
\begin{tabular}{l|llllllll|l}
\hline
Model & ar & de & es & fr & hi & it & ja & pt & \textbf{Avg.} \\
multilingual-e5-small & 21.2 & 25.6 & 26.4 & 25 & 21.1 & 26.2 & 20.7 & 24.4 & 23.8 \\
multilingual-e5-base & 23.1 & 29.8 & 29.9 & 30.9 & 22.7 & 29.8 & 22.9 & 30.6 & 27.5 \\
\textbf{granite-embedding-107m-multilingual} & 18 & 24.9 & 25.4 & 24.5 & 17.3 & 24.4 & 20.4 & 25.7 & 22.6\\
\textbf{granite-embedding-278m-multilingual} & 18.3 & 25.7 & 26 & 25.3 & 18.3 & 24.7 & 21.2 & 26.4 & 23.2 \\
\hline
\end{tabular}}
\end{adjustbox}
\caption{Mintaka Retrieval \citep{sen-etal-2022-mintaka} nDCG@10 scores.}
\label{results:mintaka}
\end{table}

We also evaluate the multilingual embedding models on multilingual retrieval tasks from the MTEB library. For brevity, we only include the per-language average scores in Table \ref{results:mteb-multi-avg}, and include detailed results on all tasks in Appendix \ref{app:multi-mteb}. As shown, the Granite Models show comparable performance with respect to other open source models, except on Chinese. This is because we do not include the in-domain training data from DuReader \citep{qiu2022dureaderretrieval} or T2Retrieval \citep{xie_2023_t2retrieval}, which are used by other multilingual models during their training.

\begin{table}[ht!]
\centering
\begin{adjustbox}{width=1\textwidth}
\setlength{\tabcolsep}{0.5em} 
{\renewcommand{\arraystretch}{1.3}
\begin{tabular}{l|llllllll}
\hline
Model & En & De & Es & Fr & Ja & Ar & Ko & Zh \\
\hline
multilingual-e5-small & 46.6 & 75.7 & 51.0 & 50.3 & 62.3 & 68.6 & 75.1 & 59.9 \\
multilingual-e5-base & 48.9 & 76.6 & 52.3 & 55.6 & 64.5 & 69.7 & 76.4 & 61.6 \\
\textbf{granite-embedding-107m-multilingual} & 45.3 & 70.3 & 48.7 & 51.1 & 59.0 & 63.2 & 70.5 & 40.8 \\
\textbf{granite-embedding-278m-multilingual} & 48.2 & 71.2 & 52.6 & 54.1 & 61.7 & 64.2 & 71.8 & 45.2 \\
\hline
\end{tabular}}
\end{adjustbox}
\caption{Multilingual MTEB Retrieval \citep{muennighoff2022mteb} nDCG@10 scores. Average scores per language shown.}
\label{results:mteb-multi-avg}
\end{table}

\subsection{Code Retrieval Performance}
We evaluate the model's code retrieval capability on the COIR benchmark \cite{li2024coircomprehensivebenchmarkcode}, consisting of 10 datasets across 7 domains in Table \ref{results:coir}. Granite Embedding models out-perform models of the same size, despite the fact that unlike other models, we do not include any code retrieval data in the training of our models, leading to purely zero-shot evaluation for our models.

\begin{table}[ht!]
\centering
\begin{adjustbox}{width=1\textwidth}
\setlength{\tabcolsep}{0.5em} 
{\renewcommand{\arraystretch}{1.3}
\begin{tabular}{p{3.2cm}|llp{1cm}p{0.8cm}p{0.8cm}p{0.8cm}p{0.8cm}p{1cm}p{1cm}p{1cm}|l}
\hline
Model & APPS & CosQA & SynText-2SQL & Code-Search Net & Code-Search Net CCR & Code-Trans-Contest & Code-Trans-DL & Stack Overflow QA & Feed-Back ST & Feed-Back MT & \textbf{Avg.} \\
\hline
e5-small-v2 & 4.4 & 29.7 & 51.9 & 55.5 & 51.3 & 53.3 & 30.8 & 83.5 & 71.2 & 39.8 & 47.1 \\
bge-small-en-v1.5 & 5.6 & 32.1 & 45.3 & 72.6 & 47.9 & 48.2 & 25.7 & 78.1 & 67.8 & 35.1 & 45.8 \\
e5-base-v2 & 11.5 & 32.6 & 51.4 & 62.7 & 56.9 & 62.5 & 21.9 & 87.9 & 74.5 & 41.6 & 50.3 \\
bge-base-en-v1.5 & 6.5 & 33.7 & 45.3 & 72.0 & 50.9 & 45.7 & 23.5 & 80.2 & 70.0 & 33.7 & 46.1 \\
\textbf{granite-embedding-30m-english} & 6.2 & 35.5 & 44.7 & 58.7 & 49.1 & 57.5 & 26.9 & 83.9 & 69.4 & 37.8 & 47.0 \\
\textbf{granite-embedding-125m-english} & 11.8 & 36.6 & 48.4 & 55.1 & 47.6 & 66.7 & 29.6 & 89.9 & 75.3 & 42.1 & 50.3 \\
\hline 
\hline
multilingual-e5-small	&12.0	&29.6	&46.2	&52.5	&55.7	&61.1	&32.3	&81.9	&72.7&	41.3	&48.5\\
multilingual-e5-base	&20.9	&31.1	&52.3	&46.6	&57.0	&43.1	&29.9	&85.1	&72.6&	43.2	&48.2\\
\textbf{granite-embedding-107m-multilingual} & 4.5 & 28.3 & 39.3 & 46.3 & 38.3 & 43.7 & 18.2 & 66.6 & 58.5 & 23.5 & 36.7 \\
\textbf{granite-embedding-278m-multilingual} & 6.1 & 33.5 & 47.3 & 52.9 & 43 & 60.8 & 32.1 & 78.4 & 67.7 & 31.4 & 45.3 \\
\hline
\end{tabular}}
\end{adjustbox}
\caption{COIR benchmark \citep{li2024coircomprehensivebenchmarkcode} nDCG@10 scores.}
\label{results:coir}
\end{table}

\subsection{Performance on IBM-related Benchmarks}

The Granite Embedding models were developed primarily with customer applications in mind. We thus evaluate the performance of the English Granite Embedding models on three IBM tasks: ClapNQ Retrieval \citep{clapnq_rosenthal_2025}, Red Hat Query Retrieval and Unified Search. 

\paragraph{ClapNQ Retrieval} consists of queries and passages, humanly modified from the Natural Questions \cite{kwiatkowski-etal-2019-natural-questions} dataset, selected such that the answers are long and non-contiguous; we evaluate performance on the 301 answerable queries from the test set, as described in \cite{clapnq_rosenthal_2025}.

\paragraph{Red Hat Query Retrieval} is an internal dataset that tests a models's ability to identify existing solutions to customer tickets regarding Red Hat products. This is formulated as a re-ranking task, where the model is used as a re-ranker over the top-100 passages selected with BM25. This test comprises of 466 test queries and a search corpus of 666K passages. 

\paragraph{Unified Search} test is another IBM internal benchmark based on IBM product documentation, used to evaluate the model's ability to retrieve the most relevant support document for a given user query regarding IBM software/hardware products. The test contains 402 queries, with a corpus of 1.7M documents.

We compare the performance of our models to other open English retrievers like Nomic Embeddings \citep{nussbaum2024nomic}, BGE models \citep{bge_embedding}, E5 models \cite{wang2022e5},Arctic Embeddings \citep{merrick2024arcticembedv1} and GTE models \citep{li2023gte}. As shown in Table \ref{results:ibm}, the Granite Embedding models outperform all other models in these internal tasks.

\begin{table}[t!]
\centering
\setlength{\tabcolsep}{0.5em} 
{\renewcommand{\arraystretch}{1.3}
\begin{tabular}{l|ccc|c}
\hline
Model & \multicolumn{1}{c}{\begin{tabular}[c]{@{}c@{}}Red Hat \\  (R@10)\end{tabular}} & \multicolumn{1}{c}{\begin{tabular}[c]{@{}c@{}}ClapNQ \\  (nDCG@10)\end{tabular}} & \multicolumn{1}{c|}{\begin{tabular}[c]{@{}c@{}}Unified Search \\  (MRR@5)\end{tabular}} & \textbf{Avg.} \\
\hline
e5-small-v2 & 40.0  & 60.3 & 27.2 & 42.5  \\
bge-small-en-v1.5 & 41.2 & 59.4 & 28.3 & 43.0\\
gte-small & 42.2 & 60.3 & 33.0 & 45.2 \\
snowflake-arctic-embed-s & 42.1 & 62.3 & 35.4 & 46.6\\
e5-base-v2 & 41.4 & 62.5 & 31.2 & 45.0 \\
bge-base-en-v1.5 & 41.2 & 65.3 & 28.0 & 44.8\\
gte-base-en-v1.5 & 40.2 & 65.1 & 29.1 & 44.8 \\
nomic-embed-text-v1.5 & 40.9 & 68.5 & 28.2 & 45.9 \\
snowflake-arctic-embed-m-v2.0 & 43.9 & 61.7 & 29.2 & 44.9\\
\textbf{granite-embedding-30m-english} & 44.0 & 64.8 & 34.0 & 47.6 \\
\textbf{granite-embedding-125m-english}& 44.0 & 71.4 & 30.0 & 48.4 \\
\hline
\end{tabular}}
\caption{IBM Benchmark evaluation. Reports nDCG@10 score for ClapNQ \citep{clapnq_rosenthal_2025}, recall@10 score for Redhat Search and MRR@5 score for Unified Search}
\label{results:ibm}
\end{table}

\subsection{Retrieval Time}

The latency of retrieval is an important aspect of the large-scale applications of embedding models. As shown in Table \ref{ret-time} The Granite Embedding models, specifically the smaller variants, have lower retrieval time per query compared to other open source models with the same vector size, while maintaining good performance. The result is based on evaluating an IBM internal benchmark with single A100 GPU. 

\begin{table}[ht!]
\centering
\setlength{\tabcolsep}{0.5em} 
{\renewcommand{\arraystretch}{1.3}
\begin{tabular}{lcc}
\hline
Model & Vector Size & Time Per Query (s) \\
\hline
e5-small-v2 & 384 & 0.31 \\
bge-small-en-v1.5 & 384 & 0.31 \\
e5-base-v2 & 768 & 0.67 \\
bge-base-en-v1.5 & 768 & 0.67 \\
\textbf{granite-embedding-30m-english} & 384 & 0.16 \\
\textbf{granite-embedding-125m-english} & 768 & 0.64 \\
\hline
\hline
multilingual-e5-small & 384 & 0.30 \\
multilingual-e5-base & 768 & 0.67 \\
\textbf{granite-embedding-107m-multilingual} & 384 & 0.17 \\
\textbf{granite-embedding-278m-multilingual} & 768 & 0.67 \\
\hline
\end{tabular}}
\caption{Retrieval Time per Query}
\label{ret-time}
\end{table}

\section{Conclusion}
We present the Granite Embedding Models, a family of open, lightweight embedding models curated for enterprise retrieval tasks. We release 5 models spanning two model sizes for a variety of latency budgets. We introduce English models that can be used in both dense and sparse vector applications, and multilingual models for dense vector search applications in 12 languages. We provide descriptions of the training data and training mechanism, with a focus on high quality data appropriate for enterprise use. We show our models achieve
higher scores in average on a variety of text embeddings and retrieval benchmarks, and show how our lightweight models can be used for applications which require very low retrieval times. All out models are released under an Apache 2.0 license, allowing for both commercial and research purposes. We plan to continuously release updates to these models to improve their performance and introduce relevant features.

\appendix

\section{Contributions}
\subsubsection*{Core Model Training}
Parul Awasthy, Radu Florian,  Yulong Li, Scott McCarley,  Arafat Sultan, Aashka Trivedi

\subsubsection*{Data and Evaluation}
Parul Awasthy, Mihaela Bornea, Radu Florian, Martin Franz, Bhavani Iyer, Vishwajeet Kumar, Rudra Murthy, Sara Rosenthal, Avirup Sil, Arafat Sultan, Aashka Trivedi

\subsubsection*{Retrieval Oriented Pretraining}
Vishwajeet Kumar, Rudra Murthy, Vignesh P, Aashka Trivedi

\subsubsection*{Product Management}
Abraham Daniels, Gabe Goodhart, Sukriti Sharma, Kate Soule

\subsubsection*{Technical Leadership}
David Cox, Radu Florian, Luis Lastras, Salim Roukos, Jaydeep Sen

\section{English Text Embedding Evaluation}
\label{app:eng-mteb}
We show the performance of the Granite Embedding models on the English MTEB benchmark \cite{muennighoff2022mteb} in Table \ref{results:mteb-full}. The MTEB benchmarks spans 8 tasks and 56 datasets, used to evaluate the quality of text embeddings for classification, clustering, pair classification, re-ranking, retrieval, re-ranking, sentence similarity and summarization tasks.

We provide the average score per task, using the main metric described in the paper: Accuracy for classification tasks, V-Measure for clustering tasks, Average Precision for pair classification tasks, MAP for re-ranking tasks, nDCG@10 for retrieval tasks and Spearman Correlation (based on cosine similarity) for STS and summarization tasks.

\begin{table}[ht!]
\centering
\begin{adjustbox}{width=1\textwidth}
\setlength{\tabcolsep}{0.5em} 
{\renewcommand{\arraystretch}{1.3}
\begin{tabular}{p{3.2cm}|p{1.2cm}lp{1.2cm}lllp{1.2cm}|l}
\hline
Model & Classi-fication & Clustering & Pair Classification & Re-ranking & Retrieval & STS & Summa-rization & \textbf{Avg.} \\
\hline
e5-small-v2 & 72.9 & 39.9 & 84.7 & 54.3 & 49.0 & 80.4 & 31.2 & 59.9 \\
bge-small-en-v1.5 & 74.1 & 43.8 & 84.9 & 58.4 & 51.7 & 81.6 & 30.1 & 62.2 \\
e5-base-v2 & 73.8 & 44.1 & 85.7 & 55.9 & 50.3 & 81.1 & 30.3 & 61.6 \\
bge-base-en-v1.5 & 75.5 & 45.8 & 86.6 & 58.9 & 53.3 & 82.4 & 31.1 & 63.6 \\
\textbf{granite-embedding-30m-english} & 61.2 & 43.2 & 78.9 & 58.5 & 49.1 & 76.2 & 29.4 & 57.3 \\
\textbf{granite-embedding-125m-english} & 62.0 & 45.3 & 79.6 & 60.3 & 52.3 & 73.1 & 31.8 & 58.4 \\

\hline
\hline
multilingual-e5-small & 67.0 & 37.2 & 82.7 & 52.9 & 46.6 & 79.2 & 30.0 & 57.1 \\
multilingual-e5-base & 69.8 & 39.6 & 83.6 & 55.0 & 48.9 & 80.3 & 30.2 & 59.1 \\
\textbf{granite-embedding-107m-multilingual} & 62.1 & 37.8 & 80.5 & 55.6 & 45.3 & 71.6 & 31.1 & 54.5\\
\textbf{granite-embedding-278m-multilingual} & 63.5 & 39.2 & 81.4 & 56.1 & 48.2 & 72.9 & 32.0 & 56.1 \\
\hline
\end{tabular}}
\end{adjustbox}
\caption{English MTEB Benchmark \cite{muennighoff2022mteb} scores.}
\label{results:mteb-full}
\end{table}

\section{Multilingual Text Retrieval Evaluation}
\label{app:multi-mteb}

To evaluate our multilingual embedding models on out-of-domain information retrieval, we use multilingual retrieval tasks from the MTEB repository. While MTEB provides an official benchmark for Polish, French, and Chinese, we pick the available retrieval tasks from German (Table \ref{results:mteb-german}), Spanish (Table \ref{results:mteb-spanish}), French (Table \ref{results:mteb-french}), Japanese (Table \ref{results:mteb-japanese}), Arabic (Table \ref{results:mteb-arabic}), Korean (Table \ref{results:mteb-korean}) and Chinese (Table \ref{results:mteb-chinese}). For English, we evaluate on the BEIR benchmark in Table \ref{results:beir}

\begin{table}[ht!]
\centering
\begin{adjustbox}{width=1\textwidth}
\setlength{\tabcolsep}{0.5em} 
{\renewcommand{\arraystretch}{1.3}
\begin{tabular}{l|llll|l}
\hline
 Model & QuAD-Retrieval & DPR & GovService-Retrieval & LegalQuAD & Avg. \\
\hline
multilingual-e5-small & 93.14 & 78.94 & 82.78 & 47.8 & 75.7 \\
multilingual-e5-base & 93.93 & 79.51 & 85.07 & 47.85 & 76.6 \\
\textbf{granite-embedding-107m-multilingual} & 90.7 & 77.2 & 80.3 & 32.8 & 70.3 \\
\textbf{granite-embedding-278m-multilingual} & 91.3 & 78.4 & 78.1 & 36.8 & 71.2 \\
\hline
\end{tabular}}
\end{adjustbox}
\caption{German Retrieval nDCG@10 scores}
\label{results:mteb-german}
\end{table}

\begin{table}[ht!]
\centering
\begin{adjustbox}{width=1\textwidth}
\setlength{\tabcolsep}{0.5em} 
{\renewcommand{\arraystretch}{1.3}
\begin{tabular}{l|ll|l}
\hline
Model & Passage Retrieval S2P &  Passage Retrieval S2S & \textbf{Avg.} \\
\hline
multilingual-e5-small & 38.99 & 62.99 & 51.0 \\
multilingual-e5-base & 39.14 & 65.52 & 52.3 \\
\textbf{granite-embedding-107m-multilingual} & 35.7 & 61.7 & 48.7 \\
\textbf{granite-embedding-278m-multilingual} & 39.1 & 66.1 & 52.6 \\
\hline
\end{tabular}}
\end{adjustbox}
\caption{Spanish Retrieval nDCG@10 scores}
\label{results:mteb-spanish}
\end{table}

\begin{table}[ht!]
\centering
\begin{adjustbox}{width=1\textwidth}
\setlength{\tabcolsep}{0.5em} 
{\renewcommand{\arraystretch}{1.3}
\begin{tabular}{l|p{1.5cm}p{1.5cm}p{1.5cm}p{1.5cm}p{1.5cm}|l}
\hline
Model & Alloprof Retrieval & BSARD Retrieval & Syntec Retrieval & FQuAD Retrieval & XPQA Retrieval & \textbf{Avg.} \\
\hline
multilingual-e5-small & 27.38 & 14.54 & 73.46 & 78.78 & 57.17 & 50.3 \\
multilingual-e5-base & 34.44 & 18.83 & 82.86 & 82.48 & 59.52 & 55.6 \\
\textbf{granite-embedding-107m-multilingual} & 38.9 & 14.9 & 79.7 & 69 & 53 & 51.1 \\
\textbf{granite-embedding-278m-multilingual} & 44.7 & 16.4 & 81.1 & 70.1 & 58.2 & 54.1 \\
\hline
\end{tabular}}
\end{adjustbox}
\caption{French Retrieval nDCG@10 scores}
\label{results:mteb-french}
\end{table}

\begin{table}[ht!]
\centering
\setlength{\tabcolsep}{0.5em} 
{\renewcommand{\arraystretch}{1.3}
\begin{tabular}{l|lll}
\hline
Model & JaQuAD Retrieval & JaGovFaq Retrieval & \textbf{Avg.} \\
\hline
multilingual-e5-small & 59 & 65.6 & 62.3 \\
multilingual-e5-base & 61.0 & 68 & 64.5 \\
\textbf{granite-embedding-107m-multilingual} & 56.4 & 61.5 & 59 \\
\textbf{granite-embedding-278m-multilingual} & 57.3 & 66.1 & 61.7 \\
\hline
\end{tabular}}
\caption{Japanese Retrieval nDCG@10 scores}
\label{results:mteb-japanese}
\end{table}

\begin{table}[ht!]
\centering
\setlength{\tabcolsep}{0.5em} 
{\renewcommand{\arraystretch}{1.3}
\begin{tabular}{l|l}
\hline
Model & Saddeem Questions \\
\hline
multilingual-e5-base & 69.7 \\
multilingual-e5-small & 68.6 \\
\textbf{granite-embedding-278m-multilingual} & 64.2 \\
\textbf{granite-embedding-107m-multilingual} & 63.2 \\
\hline
\end{tabular}}
\caption{Arabic Retrieval nDCG@10 scores}
\label{results:mteb-arabic}
\end{table}

\begin{table}[ht!]
\centering
\setlength{\tabcolsep}{0.5em} 
{\renewcommand{\arraystretch}{1.3}
\begin{tabular}{l|l}
\hline
Model & Ko-Starategy Qa \\
\hline
multilingual-e5-small & 75.1 \\
multilingual-e5-base & 76.4 \\
\textbf{granite-embedding-107m-multilingual} & 70.5\\
\textbf{granite-embedding-278m-multilingual} & 71.8 \\
\hline
\end{tabular}}
\caption{Korean Retrieval nDCG@10 scores}
\label{results:mteb-korean}
\end{table}

\begin{table}[ht!]
\centering
\begin{adjustbox}{width=1\textwidth}
\setlength{\tabcolsep}{0.5em} 
{\renewcommand{\arraystretch}{1.3}
\begin{tabular}{p{3.2cm}|p{1.2cm}p{1.2cm}p{1.2cm}p{1.2cm}p{1.2cm}p{1.2cm}p{1.2cm}p{1.2cm}|p{1.2cm}}
\hline
Model & CMedQa Retrieval & Covid-Retrieval & Du-Retrieval & Ecom-Retrieval & Medical-Retrieval & MS-Marcp & T2-Retrieval & Video-Retrieval & \textbf{Avg.} \\
\hline
multilingual-e5-small & 24.4 & 72.8 & 81.4 & 53.5 & 44.8 & 73.2 & 71.4 & 58 & 59.9 \\
multilingual-e5-base & 27.2 & 73.5 & 81.6 & 54 & 48.3 & 76 & 70.8 & 61.2 & 61.6 \\
\textbf{granite-embedding-107m-multilingual} & 16.1 & 59.7 & 57.6 & 22.1 & 27.3 & 58.5 & 57.7 & 27.4 & 40.8\\
\textbf{granite-embedding-278m-multilingual} & 19.6 & 58 & 64.1 & 28.1 & 34 & 62.5 & 63.9 & 31.6 & 45.2 \\
\hline
\end{tabular}}
\end{adjustbox}
\caption{Chinese Retrieval nDCG@10 scores}
\label{results:mteb-chinese}
\end{table}

\bibliography{iclr2025_conference}
\bibliographystyle{iclr2025_conference}

\end{document}